\documentclass[11pt]{article}
\usepackage{amssymb}
\newcommand{\ket}[1]{\left | #1 \right \rangle}
\newcommand{\bra}[1]{\left \langle #1 \right |}

\def\openone{\leavevmode\hbox{\small1\kern-3.8pt\normalsize1}}
\def\tr{{\rm tr}\; }

\def\cc{{\cal C}}

\newtheorem{theorem}{Theorem}
\newtheorem{definition}{Definition}
\newtheorem{lemma}{Lemma}

\newtheorem{corollary}{Corollary}

\newcommand{\proj}[1]{\ket{#1}\!\bra{#1}}

\newcommand{\inner}[2]{ \langle #1 | #2 \rangle}

\newcommand{\beq}{\begin{equation}}
\newcommand{\eeq}{\end{equation}}
\newcommand{\beqa}{\begin{eqnarray}}
\newcommand{\eeqa}{\end{eqnarray}}

\newcommand{\poly}{{\rm poly}}
\newcommand{\prob}{{\rm prob}}
\begin{document}
\begin{center}
{\LARGE\bf On the simulation of quantum circuits }\\
\bigskip
{\normalsize Richard Jozsa}\\
\bigskip
{\small\it Department of Computer Science, University of
Bristol,\\ Merchant Venturers Building, Bristol BS8 1UB U.K.} \\[4mm]
\date{today}
\end{center}

\begin{abstract} We consider recent works on
the simulation of quantum circuits using the formalism of matrix
product states and the formalism of contracting tensor networks.
We provide simplified direct proofs of many of these results,
extending an explicit class of efficiently simulable circuits (log
depth circuits with 2-qubit gates of limited range) to the
following: let $\cc$ be any poly sized quantum circuit (generally
of poly depth too) on $n$ qubits comprising 1- and 2- qubit gates
and 1-qubit measurements (with 2-qubit gates acting on arbitrary
pairs of qubit lines). For each qubit line $i$ let $D_i$ be the
number of 2-qubit gates that touch or cross the line $i$ i.e. the
number of 2-qubit gates that are applied to qubits $j,k$ with
$j\leq i\leq k$. Let $D=\max_i D_i$. Then the quantum process can
be classically simulated in time $n\poly(2^D)$. Thus if $D=O(\log
n)$ then $\cc$ may be efficiently classically simulated.
\end{abstract}
\bigskip
\section{Introduction}\label{intro}
The issue of finding nontrivial families of quantum circuits that
can be classically efficiently simulated is a fundamentally
important one for the understanding of quantum computational power
and for the design of new quantum algorithms. It is known
\cite{JL} that the absence of increasing multi-partite
entanglement in a quantum algorithm is sufficient to guarantee an
efficient simulation and that the question of efficient
simulability is closely related to the variety of different
possible mathematical formalisms for the representation of quantum
states and operations. To further study the origins quantum
computational speedup it is thus natural to try to identify
classes of circuits that can generate entanglement yet which can
also be classically efficiently simulated. Interesting such
families have been recently identified by Markov and Shi \cite{MS}
(using the notion of tree-width of a graph and the formalism of
contracting tensor networks) and by Yoran and Short \cite{YS}
(using matrix product state representations and the one-way
quantum computer formalism). The purpose of this note is to
present alternative derivations of many of their results with
proofs that are simpler and considerably more direct. We also
extend the identified families of efficiently simulable circuits
(although our extension may well be also amenable to analysis by
the techniques of \cite{MS} too). We refer to the introduction of
\cite{MS} for a comprehensive summary of existing results on the
simulation of quantum circuits, which we do not duplicate here.

We begin with a statement of our main result.

\begin{definition}\label{def1} Let $\cc$ be any poly sized quantum
circuit on $n$ qubits comprising 1- and 2- qubit gates. The {\em
reduced form} $\cc_{red}$ of $\cc$ is constructed as follows.
First we multiply together all 1-qubit gates that lie
consecutively along any single line and then multiply the result
into the following or preceding 2-qubit gate. This eliminates all
1-qubit gates from $\cc$. Next for every pair of lines $i,j$ we
consider all collections of consecutive 2-qubit gates acting on
lines $i,j$ that can be performed in sequence without any other
interposed gate. Each such collection is replaced by the single
2-qubit gate given by the product. The resulting circuit
$\cc_{red}$ is the reduced form of $\cc$. \end{definition}

An example of a circuit reduction is shown in figure 1.

\setlength{\unitlength}{1cm}
\begin{picture}(14,5)(1,1)

\put(1,2){\line(1,0){7.0}}

\put(1,3){\line(1,0){7.0}}

\put(1,4){\line(1,0){7.0}}

\put(1,5){\line(1,0){7.0}}

\put(10,2){\line(1,0){5.0}}

\put(10,3){\line(1,0){5.0}}

\put(10,4){\line(1,0){5.0}}

\put(10,5){\line(1,0){5.0}}

\put(2,4){\line(0,1){1.0}}

\put(3,2){\line(0,1){3.0}}

\put(3.8,3){\line(0,1){1.0}}

\put(4.4,3){\line(0,1){1.0}}

\put(5,3){\line(0,1){1.0}}

\put(6,2){\line(0,1){3.0}}

\put(6.6,3){\line(0,1){1.0}}

\put(7.4,2){\line(0,1){1.0}}

\put(7.4,4){\line(0,1){1.0}}

\put(11,4){\line(0,1){1.0}}

\put(12,2){\line(0,1){3.0}}

\put(13,3){\line(0,1){1.0}}

\put(14,2){\line(0,1){1.0}}

\put(14,4){\line(0,1){1.0}}

\put(3,2){\circle*{0.2}}

\put(2,4){\circle*{0.2}}

\put(2,5){\circle*{0.2}}

\put(3,5){\circle*{0.2}}

\put(3.8,3){\circle*{0.2}}

\put(3.8,4){\circle*{0.2}}

\put(4.4,3){\circle*{0.2}}

\put(4.4,4){\circle*{0.2}}

\put(5,3){\circle*{0.2}}

\put(5,4){\circle*{0.2}}

\put(4,5){\circle*{0.2}}

\put(5,5){\circle*{0.2}}

\put(5.6,4){\circle*{0.2}}

\put(6,2){\circle*{0.2}}

\put(6,5){\circle*{0.2}}

\put(6.6,3){\circle*{0.2}}

\put(6.6,4){\circle*{0.2}}

\put(7.4,2){\circle*{0.2}}

\put(7.4,3){\circle*{0.2}}

\put(7.4,4){\circle*{0.2}}

\put(7.4,5){\circle*{0.2}}

\put(11,4){\circle*{0.2}}

\put(11,5){\circle*{0.2}}

\put(12,2){\circle*{0.2}}

\put(12,5){\circle*{0.2}}

\put(13,3){\circle*{0.2}}

\put(13,4){\circle*{0.2}}

\put(14,2){\circle*{0.2}}

\put(14,3){\circle*{0.2}}

\put(14,4){\circle*{0.2}}

\put(14,5){\circle*{0.2}}

\put(4,1){A circuit $\cc$}

\put(10.5,1){The reduced form $\cc_{red}$}

\end{picture}

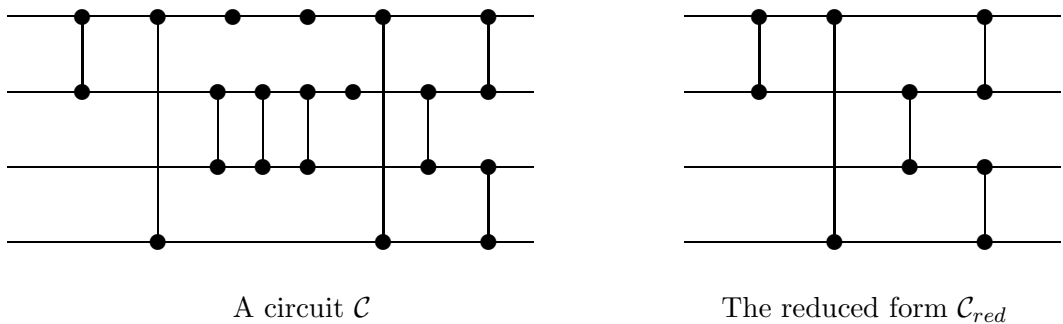
\noindent {\small Figure 1: Single black dots denote 1-qubit gates
and a dot pair connected by a vertical line denotes a 2-qubit gate
acting on the designated lines.}

It is clear that the passage from the gate description of any poly
sized circuit $\cc$ to its reduced form $\cc_{red}$ can be
calculated in poly time (as the passage involves the
multiplication of at most poly many matrices of sizes 2 by 2 or 4
by 4). Also $\cc$ and $\cc_{red}$ clearly represent the same
overall transformation. Hence $\cc$ can be classically efficiently
simulated iff $\cc_{red}$ can. Our main result is (two elementary
proofs of)  the following.

\begin{theorem} \label{myth1}
Let $\cc$  be any poly sized (and generally poly depth) quantum
circuit  on $n$ qubits comprising 1- and 2- qubit gates (with
2-qubit gates acting on arbitrary pairs of qubit lines). Let the
input be any product state of the $n$ qubits and let the output be
the result of a measurement in the standard basis on any
prescribed subset of qubits after the application of $\cc$. Let
$\cc_{red}$ be the reduced form of $\cc$ . For each qubit line $i$
in $\cc_{red}$ let $D_i$ be the number of 2-qubit gates that touch
or cross the line $i$ i.e. the number of 2-qubit gates that are
applied to qubits $j,k$ with $j\leq i\leq k$. Let $D=\max_i D_i$.
Then the output of the quantum process can be classically
simulated in time $n\poly(2^D)$. Thus if $D=O(\log n)$  then $\cc$
may be classically efficiently simulated.
\end{theorem}

We will give two direct elementary proofs of this theorem, the
first using matrix product states (MPS) and the second using just
the simple process of multiplying gates (``contracting tensors'').
Our MPS proof will extend the results of \cite{YS}. In that paper
the one way quantum computer formalism (1WQC) is combined with MPS
to show (amongst other results) that log depth circuits involving
2-qubit gates of restricted range only, can be classically
efficiently simulated. Our result improves on this in several
ways: we use only MPS without any recourse to 1WQC in our proof
and our theorem, clearly including circuits of the above type,
includes many further circuit families of fully poly depth and
involving 2-qubit gates of unrestricted range.

We note that theorem \ref{myth1} is closely similar to proposition
5.1 of \cite{MS} where the proof involves recourse to the
labyrinthine theory of tree decompositions and tree width of
graphs and associated contraction orderings. In contrast our
second proof is transparently elementary and our theorem appears
to cover essentially all the explicitly mentioned families of
efficiently simulable circuits in \cite{MS}. Thus it would be
interesting to display a family of circuits that can be seen to be
efficiently simulable from the sophisticated formalism of tree
decompositions of graphs etc. but which is not amenable to the
elementary methods that we present below.

We note that our theorem may also be readily extended to the case
where $\cc$ allows 1-qubit measurements within the body of the
circuit, with choices of later gates and measurements depending
adaptively on earlier measurement outcomes. (In that case we apply
the definition of $D_i$ to $\cc$ rather than $\cc_{red}$). But for
clarity of exposition we state our theorem in the simpler form
above. Also all results and definitions in this paper generalise
immediately from qubits to qudits for any fixed dimension  $d$ but
again for clarity we work with qubits ($d=2$) throughout.

\section{Matrix product states}\label{mpssect}
A product state of $n$ qubits has the form
\begin{equation}\label{prod} \ket{\psi}_{1\ldots n}=\ket{\alpha}_1
\ket{\beta}_2 \ldots \ket{\kappa}_n \end{equation} which
manifestly depends on a number of parameters that grows only
linearly with $n$ (in contrast to exponentially, for general
states). This restriction on the size of the state description is
significant since it allows an efficient classical simulation of
quantum processes involving such states. In the standard basis
$\ket{\psi}$ may be written as $\sum c_{i_1\ldots i_n}
\ket{i_1}\ldots\ket{i_n}$ but now we have an exponentially large
description (the set of amplitudes) which is restricted by further
conditions viz. that $c_{i_1\ldots i_n}=a_{i_1}b_{i_2}\ldots
k_{i_n}$ for some $a_i,\ldots ,k_i$'s.

A matrix product state (MPS) of $n$ qubits is a natural
generalisation of the form eq. (\ref{prod}): we simply replace
each state $\ket{\alpha},\ldots$ by a {\em matrix} of (generally
sub-normalised) states $\ket{\alpha_{ij}}\ldots$ and form a
corresponding matrix product of the $n$ matrices:
\begin{equation}\label{mps}
\ket{\psi}=\sum_{i,j,k,l,\ldots m}
\ket{\alpha_{ij}}\ket{\beta_{jk}}\ket{\gamma_{kl}}\ldots
\ket{\kappa_{mi}}. \end{equation} The sizes of the matrices may be
freely chosen subject only to the compatibility requirement for
the formation of matrix products. Note that the first and last
indices ($i$ in eq. (\ref{mps})) are also summed. If we write the
matrices $\ket{\alpha_{ij}}$ as $A$ etc. then $\ket{\psi}= \tr
(ABC\ldots K)$. The special case of eq. (\ref{prod}) is recovered
if all the matrices are 1 by 1. If we write all states in
components in the standard basis:
\[ \ket{\alpha_{ij}}=\sum_{i_1}A_{ij}^{(i_1)}\ket{i_1},\hspace{5mm}
\ldots \hspace{5mm},
\ket{\kappa_{mi}}=\sum_{i_n}K_{mi}^{(i_n)}\ket{i_n}
\] then we get the more cumbersome expression
\begin{equation}\label{mpscpt}
 \ket{\psi}=\sum_{i,\ldots ,m,i_1,\ldots ,i_n}
A_{ij}^{(i_1)}B_{jk}^{(i_2)}\ldots
K_{mi}^{(i_n)}\ket{i_1}\ket{i_2}\ldots \ket{i_n} \end{equation}
which now involves matrices of complex numbers rather than quantum
states.

\noindent {\bf Remark on PEPS:} the definition of MPS requires a
choice of {\em one}-dimensional ordering of the qubit subsystems.
However there is an alternative description of MPS -- the
so-called ``projected entangled pairs state'' description (PEPS)
\cite{VPC,VC1} with the useful feature that it generalises in a
natural way to two and higher dimensional arrays of subsystems.
Briefly it works as follows: consider the MPS of $n$ qubits in eq.
(\ref{mpscpt}) with each matrix having size at most $L$ by $L$. We
start with a sequence of $n$ maximally entangled pairs of
$L$-level systems in states
$\ket{\lambda}=\frac{1}{\sqrt{L}}\sum_{i=0}^{L-1} \ket{i}\ket{i}$
arranged in a line as depicted in figure 2.

\setlength{\unitlength}{1cm}
\begin{picture}(14,2)(0,0)

\put(1,1){$\bigstar$}

\put(3,1){$\bigstar$}

\put(3.6,1){$\bigstar$}

\put(5.6,1){$\bigstar$}

\put(6.2,1){$\bigstar$}

\put(9.6,1){$\bigstar$}

\put(10.2,1){$\bigstar$}

\put(12.2,1){$\bigstar$}

\put(1.1,1.1){\line(1,0){2}}

\put(3.7,1.1){\line(1,0){2}}

\put(6.3,1.1){\line(1,0){0.8}}

\put(8.7,1.1){\line(1,0){1}}

\put(10.3,1.1){\line(1,0){2}}

\put(7.8,1.1){$\ldots$}

\put(1.1,1.1){\oval(1.2,0.8)[r]}

\put(3.5,1.1){\oval(1.2,0.8)}

\put(6.1,1.1){\oval(1.2,0.8)}

\put(10.1,1.1){\oval(1.2,0.8)}

\put(12.4,1.1){\oval(1.2,0.8)[l]}

\put(0.6,0.2){site $n$}

\put(3,0.2){site 1}

\put(5.6,0.2){site 2}

\put(9.4,0.2){site $n\! -\! 1$}

\put(11.8,0.2){site $n$}

\end{picture}

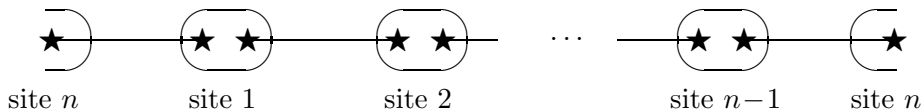
\noindent {\small Figure 2: Each line connecting a pair of stars
represents the $L$-level maximally entangled state
$\ket{\lambda}$. Each site comprises two $L$-level systems and we
apply a linear projection from $L\times L$ dimensions to 2
dimensions at each site, resulting in a state of $n$ qubits. This
PEPS is identified as having the MPS form with matrices in eq.
(\ref{mpscpt}) given directly by the matrices of the linear
projection operations.}

Consider the linear maps from $L\times L$ to 2 dimensions:
\[ P_1=A^{i_1}_{ij}\ket{i_1}\bra{ij}, \hspace{3mm}
P_2=B^{i_2}_{ij}\ket{i_2}\bra{ij}, \hspace{3mm}\ldots ,
\hspace{3mm} P_n=K^{i_n}_{ij}\ket{i_n}\bra{ij} \] applied at sites
$1,2,\ldots ,n$ respectively resulting in an $n$ qubit state
called a PEPS. Since $\ket{\lambda}$ has Kronecker delta
components $\ket{\lambda}=\frac{1}{\sqrt{L}} \sum_{ij}\delta_{ij}
\ket{i}\ket{j}$ we immediately see that the resulting PEPS of $n$
qubits is precisely the MPS in eq. (\ref{mpscpt}).

To generalise to 2 (or higher) dimensional arrays of sites we
begin instead with a 2 dimensional grid of the entangled
$\ket{\lambda}$ states. Now each site in the body of the grid has
4 $L$-level systems (or 3 or 2 at the edges or corners
respectively, of the 2-dimensional grid) and we can consider
similar site projections from each whole site into a 2-dimensional
subspace of the site. If $L$ is restricted to stay suitably small
(e.g. grow only polynomially with the number of sites) then the
resulting multi-qubit PEPS will again depend on only poly many
parameters. This formalism was introduced and exploited in
\cite{VC1} to provide ground breaking new techniques in the study
of 2 and 3 dimensional strongly correlated quantum systems in
condensed matter physics. It is also noteworthy that the
Raussendorf-Briegel cluster state (in 1 or 2 or higher dimensional
array configurations) underlying the one-way quantum computation
model is very special in having the simplest possible PEPS
description: the entangled pairs always have minimal possible
dimension $L=2$ and the projections are all the same, viz.
$P=\sum_i\ket{i}\bra{ii\ldots i}$ identifying  at each site the
qubit subspace spanned by the two kets of ``all 0's and all 1's''
(cf \cite{VC2} or \cite{Jmmt} for a more detailed description).
For applications to quantum circuits in this note we will consider
only multi-qubit states in a {\em one}-dimensional ordering.
$\Box$

It is not hard to see (cf. below for one method) that any state
$\ket{\psi}$ of $n$ qubits can be expressed in the form eq.
(\ref{mps}) but generally requiring matrices of exponential size
in $n$. Here we will be interested in those states expressible
using matrices of restricted sizes. If each matrix has size $L$ by
$L$ then the state $\ket{\psi}$ will depend on $O(nL^2)$
parameters. Hence if we restrict $L$ to be $\poly(n)$ we will
obtain a family of states (generalising product states) that
depend on only poly-many parameters, hence allowing efficient
classical simulation of their processing if we are given suitable
methods of updating the MPS description after application of gates
and measurements.

This rather abstract restriction of requiring limited matrix sizes
may be usefully related to more familiar constructs such as
Schmidt rank of bipartite divisions of $\ket{\psi}$ and
log-depthness of poly sized quantum circuits (cf. below).

To see that any state $\ket{\psi}$ of $n$ qubits may be expressed
in the MPS form we describe an explicit construction using an
iterated Schmidt decomposition, introduced by Vidal \cite{vidal}.
The qubits are always taken to be labelled $1,2,\ldots ,n$ in
linear order from left to right. We will use the following
elementary facts about Schmidt decompositions. Let
$\ket{\phi}_{AB}$ be any bipartite state with Schmidt rank $r$ and
Schmidt form
\[ \ket{\phi}_{AB}=\sum_{i=1}^r
\sqrt{\lambda_i}\ket{a_i}_A\ket{b_i}_B. \] Then:\\ {\bf Fact 1:}
if $\ket{\xi}_B$ is any state that is orthogonal to all the
$\ket{b_i}$'s then $\ket{\eta}_A\equiv \inner{\xi}{\phi}=0$.\\
{\bf Fact 2:} if $\ket{\phi}=\sum_i \ket{\xi_i}\ket{b_i}$ for some
$\ket{\xi_i}$'s then $\ket{\xi_i}=\sqrt{\lambda_i}\ket{a_i}$ for
all $i$.\\ {\bf Fact 3:} the Schmidt rank $r$ is equal to the
dimension of the support of the reduced state of $A$ in
$\ket{\phi}$.

To get an MPS form for $\ket{\psi}$ we begin by writing it in its
Schmidt form for the partition $1|2\ldots n$:
\[ \ket{\psi}=\sum_j \ket{a_j}_1\ket{\xi_j}_{2\ldots n}.\]
Here we have absorbed the Schmidt coefficients into the LH set:
$\ket{a_j}$ are orthogonal and subnormalised, and $\ket{\xi_j}$
are orthonormal. The range of the index $j$ is the Schmidt rank of
$\ket{\psi}$ for this partition. Next let $\ket{\eta_k}_{3\ldots
n}$ be the orthonormal Schmidt basis of $\ket{\psi}$ for the RH
part of the partition $12|3\ldots n$. Then for each $j$ we have
\[ \ket{\xi_j}_{23\ldots n} = \sum_k \ket{b_{jk}}_2
\ket{\eta_k}_{3\ldots n}\] where $\ket{b_k}$ are again generally
subnormalised (and non-orthogonal). To see that $\ket{\xi_j}$
cannot have a component outside the span of the $\ket{\eta_k}$'s
let $\ket{\gamma_l}$ be a set of orthonormal vectors extending
$\ket{\eta_k}$ to a full basis of the $3\ldots n$ system, so in
complete generality we can write $\ket{\xi_j}_{23\ldots n} =
\sum_k \ket{b_{jk}}_2 \ket{\eta_k}_{3\ldots n}+ \sum_{l}
\ket{c_{jl}}_2 \ket{\gamma_l}_{3\ldots n}$. Then
$\inner{\gamma_l}{\psi}=\sum_j \ket{a_j}\ket{c_{jl}}$ which must
be zero by fact 1. Since $\ket{a_j}$ are orthogonal this implies
$\ket{c_{jl}}=0$ for all $jl$.

Thus we have $\ket{\psi}=\sum_{jk}
\ket{a_j}_1\ket{b_{jk}}_2\ket{\eta_k}_{3\ldots n}$. Continuing in
this way we get \begin{equation}\label{smps}
\ket{\psi}=\sum_{j,k,l,\ldots ,p}
\ket{a_j}_1\ket{b_{jk}}_2\ket{c_{kl}}_3 \ldots \ket{k_p}.
\end{equation}
This is of the MPS form eq. (\ref{mps}): the size of the $i^{\rm
th}$ matrix is $s$ by $t$ where $s$ (resp. $t$) is the Schmidt
rank of $\ket{\psi}$ for the partition $1\ldots (s-1)|s\ldots n$
(resp. $1\ldots s|(s+1)\ldots n$). We also have further special
conditions always satisfied by this particular MPS construction:
\\ (i) the
first (resp. last) matrix has only one row (resp. one column);\\
(ii) the last matrix is an orthonormal set of states;\\ (iii) if
we consider any partition $1\ldots i|(i+1)\ldots n$ and sum the
respective parts first:
\[ \ket{\psi}= \sum_m \left( \sum_{j\ldots l}
\ket{a_j}_1\ket{b_{jk}}_2 \ldots \ket{d_{lm}}_i\right) \left(
\sum_{n\ldots p} \ket{e_{mn}}_{(i+1)} \ldots \ket{k_p}_n\right)
\equiv \sum_m \ket{A_m}_{1\ldots i}\ket{B_m}_{(i+1)\ldots n} \]
then $\{ \ket{B_m}\}$ is the orthonormal Schmidt basis for the
system $(i+1)\ldots n$ and $\{ \ket{A_m} \}$ is the orthogonal
Schmidt set for the system $1\ldots i$, subnormalised to the
corresponding Schmidt coefficients. (To see this we just halt the
iterative process leading to eq. (\ref{smps}) at stage $i$,
showing that $\{ \ket{B_m} \}$ is the orthonormal Schmidt basis
for the system $(i+1)\ldots n$ and then use fact 2).

\section{Simulating computations using MPS's}\label{simulMPS}

For any $n$ qubit state $\ket{\psi}$ let $\chi_\psi$ be the
maximal Schmidt rank of any partition $1\ldots i|(i+1)\ldots n$ of
the linearly ordered qubits into a left and right part.

Vidal\cite{vidal} and Yoran and Short\cite{YS} have shown the
following.

\begin{lemma}\label{lemma1} \cite{vidal,YS} If a single qubit
unitary gate is applied to any qubit of $\ket{\psi}$ then the MPS
description eq. (\ref{smps}) can be updated in $O(\chi_\psi^2)$
computational steps.\end{lemma}

\begin{lemma} \label{lemma2} \cite{vidal} If a 2-qubit unitary
gate is applied to any adjacent qubits (numbered $i, i+1$) of
$\ket{\psi}$ then the MPS description eq. (\ref{smps}) can be
updated in $O(\chi_\psi^3)$ computational steps.\end{lemma}

\noindent {\bf Remark:} A 2-qubit gate $U$ applied to {\em
non}-adjacent qubits on lines $l$ and $l+r$ can be replaced by
$(2r-1)$ adjacent gates viz. $(r-1)$ swaps on adjacent qubits to
make lines $l$ and $l+r$ adjacent, $U$ on adjacent qubits and
$(r-1)$ further adjacent swaps to return the lines to their
original positions, as shown in figure 3. Hence the whole process
can be simulated with $O(\chi_1^3)+\ldots +O(\chi_{2r-1}^3)$
computational cost where $\chi_i$ is $\chi$ of the $i^{\rm th}$
state in this process. We will see later (cf lemma \ref{lemma4}
and theorem \ref{myth1}) that $\max_i \chi_i$ is in fact
$O(\chi_\psi)$ (where $\ket{\psi}$ is the initial state to which
$U$ was applied) so the cost is $O(r\chi_\psi)$. In a quantum
circuit on $n$ qubits we have $r=O(n)$ so the total cost will be
$\poly(n)$ if $\chi_\psi=\poly(n)$. $\Box$

\setlength{\unitlength}{1cm}
\begin{picture}(14,5)(0,0)
\put(2,1){\line(1,0){2.0}}

\put(2,2){\line(1,0){2.0}}

\put(2,3){\line(1,0){2.0}}

\put(3,1){\line(0,1){3.0}}

\put(7,1){\line(1,0){6.0}}

\put(7,2){\line(1,0){6.0}}

\put(7,3){\line(1,0){6.0}}

\put(7,4){\line(1,0){6.0}}

\put(2,4){\line(1,0){2.0}}

\put(10,1){\line(0,1){1.0}}

\put(8,3){\vector(0,1){1}}

\put(9,2){\vector(0,1){1}}

\put(11,2){\vector(0,1){1}}

\put(12,3){\vector(0,1){1}}

\put(8,4){\vector(0,-1){1}}

\put(9,3){\vector(0,-1){1}}

\put(11,3){\vector(0,-1){1}}

\put(12,4){\vector(0,-1){1}}

\put(4.5,2.2){equals}

\put(3,1){\circle*{0.2}}

\put(3,4){\circle*{0.2}}

\put(10,1){\circle*{0.2}}

\put(10,2){\circle*{0.2}}

\put(0.6,3.9){line $l$:}

\put(0.0,0.9){line $l+r$:}

\end{picture}

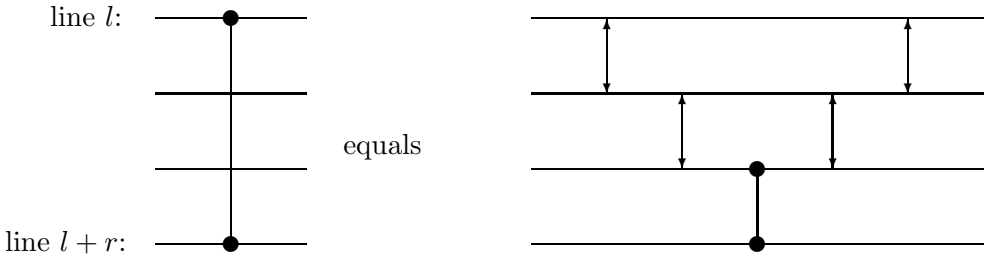
\noindent {\small Figure 3: A non-adjacent 2-qubit gate acting on
lines $r$ apart is replaced by $2r-1$ adjacent gates. The double
headed arrows denote swap gates. In this replacement each qubit
line is affected by at most 4 gates.}

\begin{lemma} \label{lemma3} \cite{YS} If a single qubit
measurement (in any chosen basis) is made on a qubit of
$\ket{\psi}$ then the outcome probabilities and MPS description
(as in eq. (\ref{smps})) of any post-measurement state can be
calculated in $\poly(\chi_\psi)$ computational steps.\end{lemma}

Vidal \cite{vidal} concluded the following: consider any pure
state poly time quantum computation with input size $n$. If the
state at every stage has $\chi$ bounded by $\poly(n)$ then the
quantum computation can be efficiently classically simulated. But
he did not relate his $\chi$-condition to any prospective
structural property of a quantum circuit.

Yoran and Short \cite{YS} noted further that the 1WQC cluster
state based on a 2 dimensional grid of size $M\times N$ has an MPS
description of the form eq.(\ref{smps}) with $\chi\leq 2^{\min
(M,N)}$. Consequently any 1WQC process (defined by a sequence of
at most $MN$ adaptive 1-qubit measurements on the $M\times N$
cluster state) can be simulated classically in $MN\poly(\chi)$
time where $\chi =2^{\min (M,N)}$. They concluded the following
result: {\em let $\cc$ be any quantum gate array on $n$ qubits
comprising 1- and 2-qubit gates such that (a) $\cc$ has log depth
and (b) the range of each 2-qubit gate is bounded by a constant
$r$ i.e. each 2-qubit gate acts on a pair of qubits at most $r$
lines apart. Then the computation can be classically efficiently
simulated. }

Their proof proceeds by noting that any such computation can be
translated by standard methods into the 1WQC formalism using a
cluster state of size $M\times N$ where $\min (M,N)= O(\log n)$
(because the circuits are log depth)
 and also $\max (M,N)=\poly (n)$.
Hence each state in the 1WQC process will have $\chi=\poly(n)$ and
lemma \ref{lemma3} gives the result.

In the translation into the 1WQC formalism, each 2-qubit gate
acting on lines $r$ apart requires a piece of cluster state of
size $O(r)\times O(r)$. If $r$ is constant, the total cluster
state for the whole (log depth) circuit will thus have a log sized
minimum dimension, but if $r$ is even allowed to be $O(\log n)$
large, then the resulting required cluster may have minimum
dimension $O((\log n)^2)$ and hence the simulation by the method
of \cite{YS} will now require $\poly (2^{O((\log
n)^2)}=\poly(n^{\log n})$ time classically.

We now introduce a further lemma about the Schmidt MPS form eq.
(\ref{smps}). This leads to a more direct proof of the above
result for quantum circuits satisfying (a) and (b)  without
recourse to the 1WQC model or cluster states. Indeed any such
circuit clearly has $D=O(\log n)$. At the same time, by proving
our theorem \ref{myth1}, we will extend the class of quantum
circuits that can be classically efficiently simulated.

\begin{lemma}\label{lemma4} Let $\ket{\psi}_{1\ldots n}$ have
Schmidt rank $r$ for the partition $A|B= 1\ldots i|(i+1)\ldots
n$.\\
(i) Let $\ket{\psi'}$ be obtained from $\ket{\psi}$ by applying a
2-qubit gate $U$ to two qubits numbered $k,l$ of $\ket{\psi}$. Let
$r'$ be the $A|B$ Schmidt rank of $\ket{\psi'}$. If $k,l$ are both
in $A$ or both in $B$ then $r'=r$. If one of $k,l$ is in $A$ and
the other in $B$ then $r'\leq 4r$.\\ (ii) Let $\ket{\psi'}$ be
obtained from $\ket{\psi}$ by application of a 1-qubit gate. Then
$r'=r$.\\ (iii) Let $\ket{\psi'}$ be any post-measurement state
resulting from a 1-qubit measurement on $\ket{\psi}$. Then $r'\leq
r$.\end{lemma}

\noindent {\bf Remark:} the bound $r'\leq 4r$ in (i) is tight. Let
$\ket{\phi^+}=\frac{1}{\sqrt{2}}(\ket{00}+\ket{11})$ and take
$\ket{\psi}_{1234}=\ket{\phi^+}_{12}\ket{\phi^+}_{34}$ with
partition $A|B=12|34$. Consider the 2-qubit gate $U$ of swap on
2,3. Then $\ket{\psi}$ has $r=1$ but after $U$ we have $r'=4$.
(For qudits the tight upper bound is a multiplicative factor of
$d^2$).

\noindent {\bf Proof of lemma \ref{lemma4}:} (i) If $k,l$ are both
on the same side of the partition then clearly $r'=r$. Thus
suppose $k,l$ lie on the two sides. Without loss of generality we
may assume that $k=i$ and $l=i+1$ since swap operations within $A$
and $B$ do not change Schmidt rank. Let $\ket{\psi}=\sum_{k=1}^r
\ket{a_k}_A\ket{b_k}_B$ be the Schmidt form. For each $k$ let
$\sigma_k$ be the reduced state of qubit $i+1$ in $\ket{b_k}$ and
let $\ket{s_k},\ket{t_k}$ be its two eigenstates. Hence in
$U\ket{\psi}$ the reduced state of $A\cup \{ i+1\}$ is spanned by
the $2r$ states $U(\ket{a_k}\ket{s_k}), U(\ket{a_k}\ket{t_k})$ for
$k=1,\ldots ,r$. Tracing out qubit $i+1$ from each of these
(generally entangled) states gives a rank 2 state of $A$ for each
$k$, so the reduced state of $A$ in $U\ket{\psi}$ is supported on
dimension at most $2(2r)$. Hence $r'\leq 4r$ follows by fact 3.

(ii) is clear. (iii) follows by noting that any post-measurement
state is obtained by applying a suitable projection to
$\ket{\psi}$. Hence the support of any reduced state cannot
increase and fact 3 gives the result.  $\Box$

\noindent {\bf Proof of theorem \ref{myth1} using MPS formalism:}
Let $\cc$ be any poly sized quantum circuit on $n$ qubits, of the
kind in the statement of the theorem. Replace each 2-qubit gate
$U$ of $\cc_{red}$ that acts on non-adjacent qubits $r$ lines
apart, by $r-1$ adjacent swap gates, $U$ on adjacent qubits, and
$r-1$ further adjacent swap gates to restore the line positions.
It is clear from figure 3 that this sequence of $2r-1$ 2-qubit
gates (each now acting on adjacent qubits) touches any given qubit
line at most 4 times (and crosses no qubit lines because of
adjacency). Thus as a result of this replacement the $D$ value
$D'$ of the circuit is increased by at most a factor of 4 ($D'\leq
4D$) and now all 2-qubit gates act on adjacent qubits. The
starting state has $\chi_{\psi_0}=1$. Consider simulating the
circuit operations in order using lemmas
\ref{lemma1},\ref{lemma2},\ref{lemma3}. Let $\ket{\psi_k}$ be the
state at any stage of the process. For any partition $1\ldots
i|(i+1)\ldots n$, $\ket{\psi_k}$ will, by lemma \ref{lemma4}, have
a Schmidt rank of at most $4^{D_i'}$ where $D_i'$ is the number of
2-qubit gates acting on lines $i,i+1$. Hence the maximal Schmidt
rank $\chi_{max}$ of any state $k$ in the process, across any
partition, is at most $4^{D'}=4^{O(D)}$. Lemmas
\ref{lemma1},\ref{lemma2},\ref{lemma3} then show that each step
can be classically simulated in $\poly(4^{D'})=\poly(4^D)$
classical computational steps and the whole process in time
$T\poly (4^D)$ where $T$ is the total number of gates in
$\cc_{red}$. Note that for any circuit $\cc$ on $n$ qubits, if $D$
is prescribed then $\cc$ can have at most $nD/2$ 2-qubit gates as
each such gate touches or crosses at least 2 lines. Hence the full
simulation time is $n\poly (4^D)=n\poly (2^D)$, and if $D=O(\log
n)$ then the simulation is efficient. $\Box$

\noindent {\bf Remark:} the $D=O(\log n)$ condition in theorem
\ref{myth1} does not imply an efficient simulation of {\em all}
poly sized log depth quantum circuits. Recall that a general poly
sized log depth quantum circuit on $n$ qubits is one for which the
gates can be transversally partitioned into $O(\log n)$ layers of
gates such that all gates in each layer can be done simultaneously
in parallel. Thus a layer may contain $O(n)$ gates so $D$ could be
$O(n)$ and hence not efficiently simulable by the method in the
proof of theorem \ref{myth1}. (For example the circuit could have
$O(n)$ gates applied to qubits
$(1,\frac{n}{2}+1),(2,\frac{n}{2}+2), \ldots , (\frac{n}{2},n)$ in
a single layer and then $D_{n/2}$ would be $O(n)$). Indeed in
\cite{TdiV} it was shown that if circuits of even {\em constant}
depth 3 (followed by a measurement layer) are efficiently
simulable then all quantum computation would be efficiently
simulable (i.e. then BQP$=$BPP). They also showed by an elementary
argument that any circuit of depth 2 (followed by a measurement
layer) {\em is} efficiently simulable. From the notions in theorem
\ref{myth1} this can be seen as follows: firstly we can reorder
the qubit lines so that all 2-qubit gates in layer 1 act on
adjacent qubits i.e. on line pairs $(1,2), (3,4),\ldots ,(n-1,n)$.
Then the gates in layer 2 may still have  $O(n)$ range but it is
straightforward to see that the line {\em pairs} can now be
reordered (thus preserving adjacency of layer 1 gates) so that
layer 2 gates have range at most 4, and the efficient simulability
then follows from theorem \ref{myth1}. $\Box$

\noindent {\bf Remark:} The condition $D=O(\log n)$ in theorem
\ref{myth1} (for efficient simulability) does not require that the
circuit be of log depth. For example a ``ladder circuit'' of
$O(n)$ 2-qubit gates applied in order to qubits $(1,2),(2,3),
\ldots ,(n-1,n)$, has $D=2$ and it is poly sized with poly depth
too. This circuit can be efficiently simulated by the method in
the proof of theorem \ref{myth1} but not by the method of ref
\cite{YS} as its 1WQC translation requires a cluster state of poly
by poly size in 2 dimensions. $\Box$

\begin{corollary}\label{cor2} (This reproduces a result from
\cite{YS}). Consider any 1WQC process on a 2 dimensional cluster
state of size $M=\poly(n)$ by $N=O(\log n)$. Then the process can
be simulated in $\poly(n)$ classical time.\end{corollary}

\noindent {\bf Proof:} Using the linear labelling used in
\cite{YS} of qubits in a cluster state of size $M\times N$ with
$N=O(\log n)$, it is clear that this cluster state can be
manufactured by a poly sized circuit with $D=O(\log n)$ and then
by theorem \ref{myth1} (or more precisely, a straightforward
extension allowing adaptive gates and measurements) the subsequent
measurement sequence can be simulated in $\poly(n)$ time too (as
1-qubit gates or measurements do not change the value of $D$).\,\,
$\Box$

\section{Contracting linear networks}\label{tensor}

We now give a completely different proof of theorem \ref{myth1}
based on the idea of ``contracting tensor networks''. This really
just amounts to multiplying out the matrices corresponding to
gates in a circuit and noting conditions under which this
calculation (with matrices of potentially exponentially growing
size) can remain only poly sized. This subject was recently
introduced into the study of quantum circuits by Markov and Shi
\cite{MS}. Our treatment here has the advantage of being much
simpler but may lack the full generality of their results.

In the above proof of theorem \ref{myth1} we made much use of
unitarity and orthogonality, for example in the very concept of a
Schmidt decomposition and unitarity preserving orthogonality in
updating the Schmidt MPS description. Thus it is surprising to
note that the tensor network approach below, based on general
linear algebraic properties only, makes no use of unitarity at all
and remains valid for arbitrary linear gates!

\noindent {\bf Proof of theorem \ref{myth1} using linear network
formalism:} Let $\cc$ be any circuit of the kind in theorem
\ref{myth1} with $D=\max D_i$. Consider first the case that the
output of the circuit is a single 1-qubit measurement on any
single qubit line, without loss of generality (wlog) the first
line.

Also assume wlog that $\cc$ has been reduced. Since $\cc$ is
poly-sized  this can be effected in $\poly(n)$ time and the
circuit now comprises only 2-qubit gates.

Furthermore assume wlog that all 2-qubit gates act on {\em
adjacent} qubit lines -- using the construction of figure 3 this
can be arranged subject only to at most a constant factor 4
increase in the value of $D$.

Suppose also wlog, that the input state for the circuit is the
$n$-qubit state $\ket{0}\ldots \ket{0}$. (Any other product input
state can then be manufactured first using only 1-qubit gates).
Let $b_i$ denote the components of the vector $\ket{0}$ in the
standard basis.

On the circuit diagram of $\cc$, for each qubit line we place an
index label on each segment between the occurrence of two 2-qubit
gates and we label the beginning of each line with the input state
$b$. This is illustrated in figure 4 for a simple typical circuit.

\begin{picture}(14,5)(-1,0)
\put(2,1){\line(1,0){3.8}}

\put(6.4,1){\line(1,0){1.6}}

\put(2,2){\line(1,0){2.4}}

\put(5,2){\line(1,0){0.8}}

\put(6.4,2){\line(1,0){1.6}}

\put(2,3){\line(1,0){1.0}}

\put(3.6,3){\line(1,0){0.8}}

\put(5,3){\line(1,0){0.8}}

\put(6.4,3){\line(1,0){1.6}}

\put(2,4){\line(1,0){1.0}}

\put(3.6,4){\line(1,0){2.2}}

\put(6.4,4){\line(1,0){1.6}}

\put(4.4,1.8){\framebox(0.6,1.4){$V$}}

\put(3,2.8){\framebox(0.6,1.4){$U$}}

\put(5.8,0.8){\framebox(0.6,1.4){$X$}}

\put(5.8,2.8){\framebox(0.6,1.4){$W$}}

\put(1.7,1){\circle{0.6}}

\put(1.7,2){\circle{0.6}}

\put(1.7,3){\circle{0.6}}

\put(1.7,4){\circle{0.6}}

\put(1.6,0.9){$b$}

\put(1.6,1.9){$b$}

\put(1.6,2.9){$b$}

\put(1.6,3.9){$b$}

\put(2.4,3.1){$j_1$}

\put(2.4,4.1){$i_1$}

\put(3,2.1){$k_1$}

\put(4,1.1){$l_1$}

\put(3.9,3.1){$j_2$}

\put(4.6,4.1){$i_2$}

\put(5.2,2.1){$k_2$}

\put(5.2,3.1){$j_3$}

\put(6.8,1.1){$l_2$}

\put(6.8,2.1){$k_3$}

\put(6.8,3.1){$j_4$}

\put(6.8,4.1){$i_3$}

\end{picture}

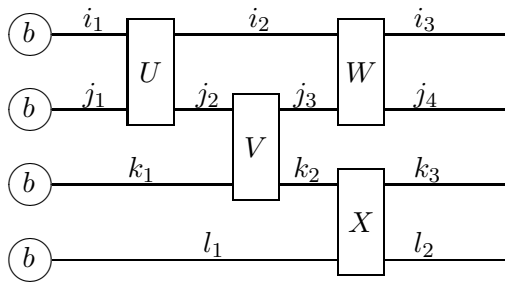
\noindent {\small Figure 4: Index labels for a simple illustrative
circuit of four 2-qubit gates $U,V,W,X$. On the first line we use
$i$'s, on the second line, $j$'s etc. for the index names. The
number of indices on any line is $O(D)$. }

Each 2-qubit gate now has 2 input indices and 2 output indices. We
write inputs as subscripts and outputs as superscripts so for
example in figure 4, $V$ has components $V_{j_2k_1}^{j_3k_2}$.
Summing over common indices corresponds to composition of the
gates in the circuit. For example in figure 4, the 4-qubit output
state $A$ has components labelled by indices $i_3,j_4,k_3,l_2$ and
is given by the contracted expression (easily read off from figure
4):
\[
A^{i_3j_4k_3l_2}=b^{i_1}b^{j_1}b^{k_1}b^{l_1}U_{i_1j_1}^{i_2j_2}
V_{j_2k_1}^{j_3k_2}W_{i_2j_3}^{i_3j_4}X_{l_1k_2}^{l_2k_3}. \] Here
all the repeated indices are understood as being summed (i.e.
contracted: the RHS has an implied
$\sum_{i_1j_1k_1l_1i_2j_2k_2j_3}$).

Next suppose we want to compute $\prob (k)$, the probability of
obtaining $k=0$ or $k=1$ from a standard basis measurement on line
1. Let $\Pi_m^n $ be the matrix of the projector $\proj{k}$. Then
for the example of figure 4, $\prob (k)$ is given by the number
obtained from the full contraction of all indices in figure 5:

\begin{picture}(14,5)(1.5,0)
\put(2,1){\line(1,0){3.8}}

\put(6.4,1){\line(1,0){3.2}}

\put(2,2){\line(1,0){2.4}}

\put(5,2){\line(1,0){0.8}}

\put(6.4,2){\line(1,0){3.2}}

\put(2,3){\line(1,0){1.0}}

\put(3.6,3){\line(1,0){0.8}}

\put(5,3){\line(1,0){0.8}}

\put(6.4,3){\line(1,0){3.2}}

\put(2,4){\line(1,0){1.0}}

\put(3.6,4){\line(1,0){2.2}}

\put(6.4,4){\line(1,0){1.2}}

\put(8.4,4){\line(1,0){1.2}}

\put(10.2,1){\line(1,0){3.8}}

\put(10.2,2){\line(1,0){0.8}}

\put(10.2,3){\line(1,0){0.8}}

\put(10.2,4){\line(1,0){2.2}}

\put(11.6,2){\line(1,0){2.4}}

\put(11.6,3){\line(1,0){0.8}}

\put(13,3){\line(1,0){1.0}}

\put(13,4){\line(1,0){1.0}}

\put(14.3,1){\circle{0.6}}

\put(14.3,2){\circle{0.6}}

\put(14.3,3){\circle{0.6}}

\put(14.3,4){\circle{0.6}}

\put(14.2,0.85){$b^\dagger$}

\put(14.2,1.85){$b^\dagger$}

\put(14.2,2.85){$b^\dagger$}

\put(14.2,3.85){$b^\dagger$}

\put(11,1.8){\framebox(0.6,1.4){$V^\dagger$}}

\put(12.4,2.8){\framebox(0.6,1.4){$U^\dagger$}}

\put(9.6,0.8){\framebox(0.6,1.4){$X^\dagger$}}

\put(9.6,2.8){\framebox(0.6,1.4){$W^\dagger$}}

\put(7.6,3.6){\framebox(0.8,0.8){$\Pi $}}

\put(12,1.1){$l_3$}

\put(13,2.1){$k_5$}

\put(10.4,2.1){$k_4$}

\put(10.4,3.1){$j_5$}

\put(11.8,3.1){$j_6$}

\put(13.2,3.1){$j_7$}

\put(9,4.1){$i_4$}

\put(11.2,4.1){$i_5$}

\put(13.2,4.1){$i_6$}

\put(4.4,1.8){\framebox(0.6,1.4){$V$}}

\put(3,2.8){\framebox(0.6,1.4){$U$}}

\put(5.8,0.8){\framebox(0.6,1.4){$X$}}

\put(5.8,2.8){\framebox(0.6,1.4){$W$}}

\put(1.7,1){\circle{0.6}}

\put(1.7,2){\circle{0.6}}

\put(1.7,3){\circle{0.6}}

\put(1.7,4){\circle{0.6}}

\put(1.6,0.9){$b$}

\put(1.6,1.9){$b$}

\put(1.6,2.9){$b$}

\put(1.6,3.9){$b$}

\put(2.4,3.1){$j_1$}

\put(2.4,4.1){$i_1$}

\put(3,2.1){$k_1$}

\put(4,1.1){$l_1$}

\put(3.9,3.1){$j_2$}

\put(4.6,4.1){$i_2$}

\put(5.2,2.1){$k_2$}

\put(5.2,3.1){$j_3$}

\put(7.9,1.1){$l_2$}

\put(7.9,2.1){$k_3$}

\put(7.9,3.1){$j_4$}

\put(6.8,4.1){$i_3$}

\end{picture}

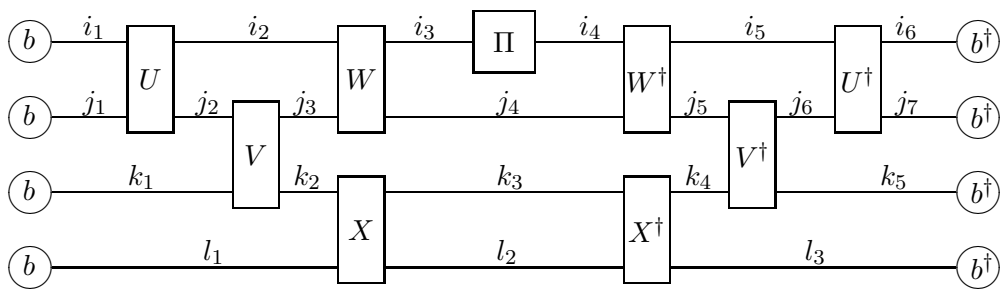
\noindent {\small Figure 5: The fully contracted linear network
whose value is $\prob (k)$. Here $\Pi$ is the 1-qubit operation
$\proj{k}$ and dagger denotes the adjoint matrix. This is the
circuit of figure 3 extended with a reflected adjoint copy and
$\Pi$ inserted in the centre on line 1.}

For a general $\cc$ (with $D=\max_i D_i$) we see that the
corresponding construction (obtained by extending $\cc$ with a
reflected adjoint copy and inserting $\Pi$ in the centre on line
1) has at most $2D+1=O(D)$ gates affecting any full single line.
It now follows that the number $\prob (k)$ can be readily
explicitly computed in $n\poly (2^{O(D)})$ steps. To do this we
consider the full tensor expression (as depicted in figure 5 for
our illustrative example) and sum all indices on line 1 e.g. all
$i$-indices in figure 5. There are $O(D)$ $i$-indices each taking
values 0 and 1 i.e. a total of $2^{O(D)}$ terms in the sum for
each set of $O(D)$ $j$-index values. These $j$-indices have
$2^{O(D)}$ sets of values too so the total computational effort to
do all the corresponding $i$-sums, for all $j$-values, is
$2^{O(D)}\poly (2^{O(D)})=\poly (2^{O(D)})$.

Having summed the $i$-indices we are left with a single object
with $O(D)$ $j$-indices and further 2-qubit gates with indices
$j,k,\ldots ,l$.

Next we sum out all the $j$-indices. This again requires
$2^{O(D)}$ sums (for all the $k$-index values) each of which is a
sum over $2^{O(D)}$ terms (i.e. the sum over all $j$-index
values). Continuing in this way, if there are $n$ qubit lines the
final number $\prob (k)$ is computed in $n \poly (2^{O(D)})$
steps. If $D=O(\log n)$ this whole computation is poly time.

In this way we can compute $\prob (k)$ for $k=0,1$ and then sample
the resulting distribution. This completes the simulation of the
quantum qubit measurement outcome.

Next suppose the output is not just a single measurement but the
measurement of a subset $\{ a,b,\ldots ,e \}$ of the qubit lines
(a subset whose size could be $O(n)$). To simulate this, we first
compute as above, the distribution for line $a$ only and sample
the distribution to get an outcome $k_a$ say. Then we place the
matrix $\Pi (k_a)/\sqrt{\prob (k_a)}$ on line $a$ and calculate
the distribution for line $b$ given that line $a$ has value $k_a$,
by repeating the above procedure. Then we sample the resulting
$\prob (k_b|k_a)$ distribution to get a value $k_b$ for line $b$.
Continuing in this way we sample the whole required joint
distribution on lines $a,b,\ldots ,e$. If $D=O(\log n)$ this whole
simulation is clearly still poly time as the size of the set of
lines is at most $O(n)$.

Finally consider the more general type of circuit $\cc$ in which
1-qubit measurements can be performed in the body of the circuit
and choice of later gates may depend on previous measurement
outcomes. To simulate this situation we consider the circuit $\cc$
only up to its first measurement and simulate the output
distribution as above. After sampling it we place the matrix of
the corresponding projector, divided by its square-root
probability, on its line and fix the identity of any later gates
that depended on this outcome. Continuing in this way with each
subsequent measurement in order of occurrence we simulate the
whole circuit $\cc$. Since $\cc$ has at most $\poly (n)$ such
measurements this entire simulation will be poly time if $D=O(\log
n)$.

\bigskip
\noindent {\large\bf Acknowledgements}\\ This work was supported
in part by the UK's EPSRC-QIPIRC network and by the European
Commission under the Integrated Project Qubit Applications (QAP)
funded by the IST directorate as Contract Number 015848.

\end{document}